\documentstyle[12pt,epsfig]{article}
\def\fe#1#2{
\begin{picture}(100,200)(0,0)
\put(34,0){\circle*{5}}
\put(0,35){\vector(1,-1){35}}
\put(-2,26){\makebox(0,0){#1}}
\put(-2,-26){\makebox(0,0){#2}}
\put(0,-35){\vector(1,1){35}}
\end{picture} }
\def\bp#1#2{
\begin{picture}(100,200)(0,0)
\multiput(0,0)(16,0){3}{\oval(8,8)[t]}
\multiput(8,0)(16,0){3}{\oval(8,8)[b]}
\put(#2,15){\makebox(0,0){#1}}
\end{picture}   }
\def\fpe#1{
\begin{picture}(100,200)(0,0)
\put(0,0){\line(1,0){35}}
\put(5,10){\makebox(0,0){#1}}
\end{picture}   }
\def\ft#1{
\begin{picture}(100,200)(0,0)
\put(0,0){\line(0,1){68}}
\put(10,24){\makebox(0,0){#1}}
\end{picture}   }
\def\hpe#1#2{
\begin{picture}(100,200)(0,0)
\multiput(0,0)(10.8,0){5}{\line(1,0){5.2}}
\put(#2,15){\makebox(0,0){#1}}
\end{picture}   }

\def\fsp#1#2{
\begin{picture}(100,200)(0,0)
\put(0,0){\circle*{5}}
\multiput(0,0)(16,16){3}{\line(1,1){10}}
\multiput(0,0)(16,-16){3}{\line(1,-1){10}}
\put(50,36){\makebox(0,0){#1}}
\put(52,-36){\makebox(0,0){#2}}
\end{picture}   }
\def\fspp#1{
\begin{picture}(100,200)(0,0)
\put(0,0){\circle*{5}}
\multiput(0,0)(16,16){3}{\line(1,1){10}}
\put(50,36){\makebox(0,0){#1}}
\end{picture}   }

\def\bsbas#1{
\begin{picture}(100,200)(0,0)
\multiput(5,0)(8,-8){5}{\oval(8,8)[bl]}
\multiput(5,-8)(8,-8){5}{\oval(8,8)[tr]}
\put(34,-30){\makebox(0,0){#1}}
\end{picture}   }
\begin{document}
\begin{flushright}
KEK Preprint 99-107\\
UFR-HEP/99/12 \\
October 1999 \\
\end{flushright}
\vspace{1cm}
\begin{center}
{\Large\sc {\bf Charged and Pseudoscalar Higgs production
at a Muon Collider}}
\vspace*{3mm}
{\large\sc {\bf }}
\vspace{1cm}

{\large
{{ A.G. Akeroyd}$^{\mbox{a}}$, { A. Arhrib}$^{\mbox{b,c}}$}
and C. Dove }
\vspace{1cm}
{\sl

a: KEK Theory Group, Tsukuba,\\
 Ibaraki 305-0801, Japan\\

\vspace{1cm}

b: D\'epartement de Math\'ematiques, Facult\'e des Sciences et Techniques\\
B.P 416, Tanger, Morocco

\vspace{1cm}

c: UFR--High Energy Physics, Physics Departement, Faculty of Sciences\\
PO Box 1014, Rabat, Morocco
}

\end{center}

\vspace{2cm}
\begin{abstract}
We consider single charged Higgs ($H^{\pm}$) and pseudoscalar Higgs
($A^0$) production in association with a gauge boson
at $\mu^+\mu^-$ colliders.
We find that the tree--level t--channel and s--channel
contributions to $\mu^+\mu^-\to H^{\pm}W^{\mp}, A^0Z$
are enhanced for large
values of $\tan\beta$, allowing sizeable cross--sections
whose analogies at $e^+e^-$ colliders would be very small.
These processes provide attractive new ways of producing
such particles at $\mu^+\mu^-$ colliders and are superior
to the conventional methods in regions of parameter space.
\end{abstract}

\newpage
\section{Introduction}
Charged Higgs bosons ($H^{\pm}$) are predicted in many favourable
extensions of the
Standard Model (SM), in particular the Minimal Supersymmetric
Standard Model (MSSM). Their phenomenology \cite{Gun} has received
much attention both at $e^+e^-$ colliders \cite{e+e-}
and at hadron colliders \cite{Hadron}, \cite{recent}. It is well known that
 $e^+e^-$ colliders offer a much cleaner environment
in which to look than hadron colliders, although recently progress
has been made
in the possibilities of detecting $H^{\pm}$ for $M_{H^{\pm}}
\ge m_t$ at hadron colliders \cite{progress}.
At $e^+e^-$ colliders production proceeds via the
mechanism $e^+e^-\to \gamma^*,Z^*\to H^+H^-$,
with higher order corrections
evaluated in \cite{eeHH}, and detection is possible for
$M_{H^{\pm}}$ up to approximately $\sqrt s/2$. The combined
null--searches from all four LEP collaborations derive the lower
limit $M_{H^{\pm}}\ge 77.3$ GeV $(95\%\, c.l)$ \cite{LEP}.

In recent years an increasing amount of work has been dedicated to
the physics possibilities of $\mu^+\mu^-$ colliders
\cite{Muon}, \cite{talks}. Such colliders
offer novel ways of producing Higgs bosons, such as an $s$--channel
resonance in the case of neutral scalars.
The existing studies do not highlight any
difference between the charged Higgs phenomenology
at a $\mu^+\mu^-$ collider and $e^+e^-$ collider,
and state that the main production mechanism would be
via $\mu^+\mu^-\to \gamma^{*},Z^*\to H^+H^-$. The rate for this process is
identical at both colliders. In the MSSM $H^{\pm}$
becomes roughly degenerate in mass with $H^0$ and $A^0$ for masses greater
than 200 GeV.
It is this correlation among the masses of the Higgs bosons
which disallows any large effects from
a $s$--channel resonance (via $\mu^+\mu^-\to H^0,h^0\to H^+H^-$)
in the pair production mode, and we
explicitly confirm this.
In order for the above to be maximised one would require
$\sqrt s \approx M_{h,H}\ge 2M_{H^{\pm}}$, a condition
which requires sizeable mass splittings among the Higgs bosons and is
disallowed in the MSSM.

So far unconsidered is the process
$\mu^+\mu^-\to H^{\pm}W^{\mp}$ via
s--channel and t--channel diagrams. Na\" \i vely, this may offer
greater possibilities of a large rate since the Yukawa coupling only
appears at one vertex in contrast to both vertices in the
pair production case.
In addition, it offers the possibility of searching for $M_{H^{\pm}}$ up to
$\sqrt s-M_W$ in contrast to pair production which only probes up to
$M_{H^{\pm}}\le \sqrt s/2$. The rate
for $b\overline b\to H^{\pm}W^{\mp}$ at hadron colliders
was considered in Ref. \cite{bbwh} although is not
expected to provide an observable signature above the background
\cite{MorOda}, at least at LHC energies. In contrast,
$\mu^+\mu^-\to H^{\pm}W^{\mp}$ might give a clean signature, since
backgrounds are considerably less.

In an analogous way we also consider $\mu^+\mu^-\to A^0Z$.
The phenomenology of $A^0$ is made tricky at $e^+e^-$ colliders due to the
absence of a tree--level vertex $ZZA^0$
and so the standard Higgsstrahlung mechanism ($e^+e^-\to A^0Z$)
 only proceeds via loops \cite{eeza}.
Moreover, over a wide region of parameter space in the MSSM
$A^0$ has a suppressed rate in the
channel $\mu^+\mu^-\to A^0h^0$, while  $\mu^+\mu^-\to A^0H^0$ only probes
up to $M_{A}\approx \sqrt s/2$. Proposed search strategies at
$\mu^+\mu^-$ collider include the scanning technique and Bremsstrahlung
tail method. Since both may provide a challenge for machine and detector
design we consider the prospects of searching for $A^0$ via
$\mu^+\mu^-\to A^0Z$.

Our work is organized
as follows. In Section 2 we perform the full tree--level calculation
of $\mu^+\mu^-\to H^+H^-$,
$\mu^+\mu^-\to H^{\pm}W^{\mp}$ and $\mu^+\mu^-\to A^0Z$.
In Section 3 we present numerical values of the cross--sections
and Section 4 contains our conclusions.
\section{Calculation}
We now consider in turn the various production mechanisms. Our calculations
are valid in both the MSSM and a general Two--Higgs--Doublet--Model (2HDM),
the difference being that the MSSM Higgs sector is parametrized by just two
parameters at tree--level (usually taken as $M_A$ and $\tan\beta$), while
the 2HDM contains 7 free parameters. Thus in a general 2HDM all four Higgs
boson masses may be taken as independent, as well as the two mixing angles
$\alpha$ and $\beta$, and the Higgs potential parameter $\lambda_5$
(in the notation of Ref. \cite{Djouadi}). In addition, the Higgs trilinear
couplings differ from those in the MSSM. In this paper we shall present
numerical results for the MSSM.
Let us summarise the couplings needed for our study:\\
\\
{\bf Fermion--Fermion--Higgs couplings}
\begin{eqnarray}
& & h^0 \mu^+\mu^- =-\frac{igm_\mu}{2 M_W} \lambda_{h\mu^+\mu^-} \qquad \quad ,
\qquad \qquad  H^0 \mu^+\mu^- =-\frac{igm_\mu}{2 M_W} \lambda_{H\mu^+\mu^-}
\nonumber\\
& & A^0 \mu^+\mu^- =-\frac{ig m_\mu}{2 M_W}\gamma_5
\lambda_{A^0\mu^+\mu^-} \ \ ,
\ \  H^- \mu^+\nu_\mu =\frac{igm_\mu}{\sqrt{2} M_W}
 \lambda_{H^+\mu^+\nu_\mu}\frac{1-\gamma_5}{2}
\end{eqnarray}
In the MSSM these couplings are given by:
\begin{eqnarray}
& & \lambda_{h\mu^+\mu^-}=-\frac{\sin\alpha}{\cos\beta} \ \ , \ \
\lambda_{H\mu^+\mu^-} =\frac{\cos\alpha}{\cos\beta}     \nonumber\\ & &
\lambda_{A^0\mu^+\mu^-}=\tan\beta  \ \ , \ \
\lambda_{H^-\mu^+\nu_{\mu}}=\tan\beta
\end{eqnarray}
One can see from the above formula that the CP--odd $A^0$ and the
charged Higgs
bosons coupling to the $\mu^{\pm}$ can be enhanced for large $\tan\beta$.

The momenta of the incoming $\mu^+$ and $\mu^-$, outgoing
gauge boson $V$ ($W^{\pm}$ or $Z$)
and outgoing Higgs scalar
$S$ ($H^{\pm}$ or $A^0$) are denoted by $p_{\mu^+}$, $p_{\mu^-}$,
$p_V$ and $p_{S}$,
respectively. Neglecting the muon mass $m_{\mu}$,
the momenta in the centre of mass of the $\mu^+\mu^-$ system are given by:
\begin{eqnarray}
& & p_{\mu^-,\mu^+}=\frac{\sqrt{s}}{2} (1,0,0,\pm 1) \nonumber \\
& & p_{V,A^0}=\frac{\sqrt{s}}{2} (1\pm \frac{M_V^2 -M_{S}^2}{s},
\pm \frac{1}{s}\lambda^{\frac{1}{2}}(s,M_V^2,M_S^2) \sin\theta,0,\pm \frac{1}{s}\lambda^{\frac{1}{2}}(s,M_V^2,M_S^2)
\cos\theta), \nonumber
\end{eqnarray}
\noindent
Here $\lambda(x,y,z)=x^2+y^2+z^2-2xy-2xz-2yz$
is the two body phase space function and
$\theta$ is the scattering angle
between $\mu^+$ and $S$;
%$$\kappa ^2=(s-(M_{S} + M_V)^2)(s-(M_{S} -M_V)^2)/s^2 ;$$
$M_{V}$ is the mass  of the gauge boson $V$
and $M_{S}$ is the mass of the Higgs scalar $S$.
In the case of $H^+H^-$ production replace $V$ by $S$.
The Mandelstam variables are defined as follows:
\begin{eqnarray}
& & s =  (p_{\mu^-}+p_{\mu^+})^2 = (p_V+p_{S})^2  \nonumber\\
& & t = (p_{\mu^-}-p_V)^2 = (p_{\mu^+}-p_{S})^2 =
\frac{1}{2}(M_V^2  + M_{S}^2) -
\frac{s}{2}
+\frac{1}{2} \lambda^{\frac{1}{2}}(s,M_V^2,M_S^2)
\cos\theta  \nonumber\\
& & u = (p_{\mu^-}-p_{S})^2 = (p_{\mu^+}-p_V)^2 =
\frac{1}{2} (M_V^2 + M_{S}^2) -
\frac{s}{2}-
\frac{1}{2} \lambda^{\frac{1}{2}}(s,M_V^2,M_S^2) \cos\theta \nonumber \\
& & s+t+u = M_V^2 + M_{S}^2  \nonumber
\end{eqnarray}

\subsection{$\mu^+\mu^-\to H^+H^-$}

\noindent
%  mu mu -> H+ H-
\begin{center}
\begin{picture}(100,100)(0,0)
\put(-160,69){\vector(1,-1){1}}
\put(-160,31){\vector(-1,-1){1}}
\put(-180,50) {\fe{$\mu^-$}{$\mu^+$}}
\put(-140,50) {\bp{$\gamma,Z$}{24}}
\put(-95,50) {\fsp{$H^-$}{$H^+$} }
\put(0,69){\vector(1,-1){1}}
\put(0,31){\vector(-1,-1){1}}
\put(-20,50) {\fe{$\mu^-$}{$\mu^+$}}
\put(16,50) {\hpe{$h,H$}{24}}
\put(65,50) {\fsp{$H^-$}{$H^+$} }
\put(159,85){\vector(1,0){1}}
\put(159,15){\vector(-1,0){1}}
\put(180,50){\vector(0,-1){1}}
\put(140,85) {\fpe{$\mu^-$}}
\put(180,85){\circle*{5}}
\put(176,16) {\ft{$\ $}}
\put(165,44) {$\nu_\mu$}
\put(180,16){\circle*{5}}
\put(140,15) {\fpe{$\mu^+$}}
\put(180,85) {\hpe{$\ $}{20}}
\put(182,16) {\hpe{$\ $}{30}}
\put(215,88) {$H^-$}
\put(215,24) {$H^+$}
\put(35,-10) {{\bf Figure.1}}
\end{picture}
\end{center}
\vspace{0.6cm}
This process proceeds via the conventional Drell--Yan mechanism
$\mu^+\mu^-\to \gamma^*,Z^*\to H^+H^-$, the analogy of
$e^+e^-\to \gamma^*,Z^*\to H^+H^-$.
Since $m_{\mu}\approx 200 m_e$
one may consider the s--channel and t--channel diagrams
(see Fig.~1), whose analogies at $e^+e^-$ colliders would be suppressed
by factors of $m_e$. The s--channel diagrams would be
maximised for $\sqrt s=
M_{h}$ or $M_H$, although in the context of the MSSM this condition
would not allow on--shell pair production of $H^{\pm}$.
This can seen from the fact that
$\sqrt s \ge 2M_{H^{\pm}}$ and $\sqrt s\approx M_{h}$ or $M_H$ cannot be
simultaneously satisfied in the MSSM.
In contrast, such s--channel diagrams were considered in Ref.\cite{Porod}
for squark production via the process $\mu^+\mu^- \to
\widetilde q\widetilde q$, and were shown to cause a doubling of the
cross--section at resonance. The t--channel diagram in Fig.~1 suffers from
Yukawa coupling suppression at two vertices.
In the calculation we shall use the following notation:
\begin{eqnarray}
Y_V & = & - Y_A =  \frac{ m_\mu^2  }{4s_W^2 M_W^2} \lambda_{H^-\mu\nu_\mu}^2
\nonumber\\
 a_h &= & -\frac{ g_{h H^+ H^-}
m_\mu \lambda_{h\mu^+\mu^-}}{2  M_W s_W } \ \ \ , \ \ \
a_H  =  -\frac{g_{H H^+ H^-}
m_\mu  \lambda_{H\mu^+\mu^-}}{2 M_W s_W }\nonumber\\
a_1 & = & -
2 \frac{1}{s} - \frac{1}{2 s_W^2 c_W^2} \frac{g_H g_V}{s-M_Z^2+ i M_Z \Gamma_Z} -
\frac{Y_V}{t} \nonumber  \\
a_2 & = &
 \frac{1}{2 s_W^2 c_W^2} \frac{g_H g_A}{s-M_Z^2+ i M_Z \Gamma_Z} -
\frac{Y_A}{t}  \nonumber \\
a_3 & = &  \frac{a_h}{s-M_h^2+i M_h \Gamma_h} + \frac{a_H}{s-M_H^2+ i M_H \Gamma_H} +
\frac{m_\mu Y_V }{t}
\end{eqnarray}
where $g_V=-(1- 4 s_W^2)/2$, $g_A=-1/2$ and
$g_H= - c_W^2 + s_W^2$.
The coupling $g_{h H^+ H^-}$ and $g_{H H^+ H^-} $
(normalised to electric charge e) are given by:
\begin{eqnarray}
g_{ H  H^+H^-}& =& - \frac{1}{s_W} \{M_W \cos(\beta - \alpha) - \frac{M_Z}{2 c_W}
\cos 2 \beta \cos(\beta + \alpha)  +
 \epsilon \frac{\cos\alpha\cos^2\beta}{2 c_W M_Z \sin\beta} \}
 \nonumber \\
g_{hH^+H^-}  & =& - \frac{1}{s_W} \{ M_W \sin(\beta - \alpha) + \frac{M_Z}{2 c_W}
\cos 2 \beta \sin(\beta + \alpha)
+ \epsilon \frac{\sin\alpha\cos^2\beta}{2 c_W M_Z \sin\beta}
\} \nonumber
\end{eqnarray}
Where
\begin{equation}
\epsilon={3G_Fm^4_t\over \sqrt 2\pi^2\sin^2\beta}
\log\left[\frac{ m_{\tilde t_1}m_{\tilde t_2}}{m^2_t}\right]
\end{equation}
The $\epsilon$ term corresponds to the leading log 1--loop corrections
\cite{epsilon} to the trilinear couplings. We will include also these
leading log corrections to the Higgs--masses and to the mixing angles.

The square amplitude is given by:
\begin{eqnarray}
|M|^2 &=& e^4\{ (|a_1|^2 + |a_2|^2) \frac{s^2}{2}\beta_H^2 \sin^2\theta
-2 |a_2|^2 m_\mu^2 s \beta_H^2  +2 |a_3|^2 s \nonumber \\ & &
  + 4 \Re(a_1 a_3) m_\mu s
\beta_H \cos\theta   \}
\end{eqnarray}
with $\beta_H^2 =1 -4 M_{H^\pm}^2/s$.
The differential cross--section is given by:
\begin{eqnarray}
\frac{d\sigma}{d\Omega}=\frac{\beta_H}{64 \pi^2 s}\frac{1}{4}
|M|^2 \label{ref0}
\end{eqnarray}

\subsection{$\mu^+\mu^-\to H^{\pm}W^{\mp}$}
%%%%%%%
\noindent
%
% mu mu -> H+ W-
\par
\begin{center}
\begin{picture}(100,100)(0,0)
\put(-40,69){\vector(1,-1){1}}
\put(-40,31){\vector(-1,-1){1}}
\put(-60,50) {\fe{$\mu^-$}{$\mu^+$}}
\put(-24,50) {\hpe{$h,H,A^0$}{24}}
\put(25,50) {\fspp{$H^-$} }
\put(23,50) {\bsbas{$\ \ \qquad W^+ $}}
\put(129,90){\vector(1,0){1}}
\put(129,20){\vector(-1,0){1}}
\put(150,55){\vector(0,-1){1}}
\put(110,90) {\fpe{$\mu^-$}}
\put(150,90){\circle*{5}}
\put(146,21) {\ft{$\ $}}
\put(135,49) {$\nu_\mu$}
\put(150,21){\circle*{5}}
\put(110,20) {\fpe{$\mu^+$}}
\put(150,90) {\hpe{$\ $}{20}}
\put(152,21) {\bp{$\ $}{30}}
\put(185,93) {$H^-$}
\put(185,29) {$W^+$}
\put(42,-10) {{\bf Figure.2}}
\end{picture}
\end{center}
\vspace{0.6cm}
Single $H^{\pm}$ production may proceed via an s--channel resonance
mediated by
$h^0,H^0$ or $A^0$, and by t--channel exchange of $\nu_{\mu}$
(see Fig.~2). All are negligible at an $e^+e^-$ collider due to the
smallness of $m_e$. The loop induced contributions to
$e^+e^-\to H^{\pm}W^{\mp}$ were considered in
Ref.\cite{Chinese} and shown to reach a few fb at very
low values of $\tan\beta$, a region disfavoured in the MSSM.
Potential advantages of {$\mu^+\mu^-\to H^{\pm}W^{\mp}$ over
standard pair production are the following:

\begin{itemize}
\item
$\mu^+\mu^-\to H^{\pm}W^{\mp}$ is sensitive to the
$H^{\pm}\mu^{\mp}\nu_{\mu}$ Yukawa coupling, which is model dependent,
and hence provides information on the underlying Higgs
structure. For example, we shall see that a 2HDM with the
Model~I type structure
would not register a signal in this channel. In contrast
$\mu^+\mu^-\to \gamma^*,Z^*\to
H^+H^-$ has a model independent rate.

\item Single $H^{\pm}$ production is
less phase space suppressed than $H^{\pm}$ pair production,
and would also allow greater kinematical reach at a given collider
(on--shell production up to $\sim \sqrt s-M_W$).

\item
The t--channel contribution may be sizeable and does not require
$\sqrt s\approx M_{res}$ to be significant, where $M_{res}$ is the mass of
a neutral Higgs s--channel resonance.
This is in contrast to other novel production processes at
$\mu^+\mu^-$ colliders, which usually require the condition
$\sqrt s\approx M_{res}$.
\end{itemize}
The differential cross--section for
$\sigma(\mu^+\mu^-\to H^{\pm}W^{\mp})$ may be written as follows:
\begin{equation}
{d\sigma\over d\Omega} = {\lambda^{1\over 2}(s,M_{H^{\pm}}^2,M_W^2)\over
64\pi^2s^2} |{\cal M}|^2 \label{ref1}
\end{equation}
The matrix element squared is given by:
\begin{eqnarray}
|{\cal M}|^2 & = &  \frac{s g^4 m_{\mu}^2 }{32M_W^4}
[ (|a_V|^2+|a_A|^2) \lambda( s,M_{H^\pm}^2,M_W^2)
+ 2 a_t^2(2 M_W^2 p_T^2 + t^2 )  \nonumber\\ [0.12cm]
& & \qquad\qquad +
2a_t(M_{H^\pm}^2 M_W^2-s p_T^2-t^2)\Re(a_V-a_A) ]
\end{eqnarray}
Where $p^2_T=\lambda(s,M_{H^{\pm}}^2,M_W^2) \sin^2\theta/4s$
and the couplings $a_V$,$a_A$ and $a_t$ are given by:
\begin{eqnarray}
a_V &=& \left( {\cos(\alpha-\beta)\lambda_{h\mu^+\mu^-}\over s-M_h^2+
iM_h\Gamma_h}
+{\sin(\alpha-\beta)\lambda_{H\mu^+\mu^-}\over s-M_H^2+iM_H\Gamma_H}\right) \\
a_A &=& {\lambda_{A\mu^+\mu^-}\over s-M_A^2+iM_A\Gamma_A} \\
a_t &=& {\lambda_{H^-\mu^+\nu_{\mu}}\over t}
\end{eqnarray}
The mixing angle dependence of the Higgs--Fermion--Fermion couplings
is contained in $\lambda_{h\mu^+\mu^-}$, $\lambda_{H\mu^+\mu^-}$,
$\lambda_{A\mu^+\mu^-}$ and $\lambda_{H^-\mu^+\nu_{\mu}}$.

Our formula agrees with that for $b\overline b\to H^{\pm} W^{\mp}$
in Ref.~\cite{bbwh}, with the
replacements $m_t\to m_{\nu_{\mu}}$ and $m_b\to m_{\mu}$.
Due to CP--invariance the rate for $W^+H^-$ and $W^-H^+$ production is
identical. The total cross section takes the following form:
\begin{eqnarray}
\sigma(\mu^+\mu^-  & \to & W^+ H^-)  =  \frac{G_F m_\mu^2}{16 \pi s^2 }
 \{ ( |a_V|^2 + |a_A|^2) \lambda(s,M_{H^\pm}^2,M_W^2 )  s  \\ [0.1cm] & + &
 2 \tan\beta
 [\Re (a_A -a_V) (M_{H^\pm}^2 + M_W^2 - s) s
 + (s - 4 M_W^2 ) \tan\beta ] \lambda^{\frac{1}{2}}(s,M_{H^\pm}^2,M_W^2 )
  \nonumber \\ [0.1cm] & - &
 4 M_W^2 \tan\beta [  \Re (a_V - a_A)   M_{H^\pm}^2 s +
 (M_{H^\pm}^2 + M_W^2 - s) \tan\beta  ] F(s,M_{H^\pm}^2,M_W^2) \}\nonumber
\end{eqnarray}
with:
$$F(s,M_{S}^2,M_V^2)=Log[\frac{M_{S}^2 + M_V^2 - s - \lambda^{\frac{1}{2}}
(s,M_{S}^2,M_V^2 ) }
{ M_{S}^2 + M_V^2 - s + \lambda^{\frac{1}{2}}(s,M_{S}^2,M_V^2 ) }]$$

\subsection{$\mu^+\mu^-\to A^0Z$}
As depicted in Fig. 3, this process proceeds in a very similar way to that 
for $\mu^+\mu^-\to H^{\pm}W^{\mp}$, except there are two t--channel diagrams.
The process $\mu^+\mu^-\to Z\phi^0$, where $\phi^0$ is the SM Higgs boson,
has been considered in Ref.~\cite{Tik}. Our calculation differs since
there is no s--channel $Z$ exchange for $\mu^+\mu^-\to A^0Z$ in the MSSM.
Instead there are two s--channel Higgs exchange diagrams of similar
magnitude to the
t--channel diagram, giving rise to strong interference. In addition
$\tan\beta$ plays an important role.
In the SM the s--channel $Z$ exchange is the dominant diagram at the
collider energy we consider ($\sqrt s=500$ GeV), and
so interference is minimal.

\noindent
% mu mu -> A Z
\par
\begin{center}
\begin{picture}(100,100)(0,0)
\put(-140,69){\vector(1,-1){1}}
\put(-140,31){\vector(-1,-1){1}}
\put(-160,50) {\fe{$\mu^-$}{$\mu^+$}}
\put(-124,50) {\hpe{$h,H$}{24}}
\put(-75,50) {\fspp{$A^0$} }
\put(-77,50) {\bsbas{$\ \ \qquad Z $}}
\put(29,90){\vector(1,0){1}}
\put(29,20){\vector(-1,0){1}}
\put(50,55){\vector(0,-1){1}}
\put(10,90) {\fpe{$\mu^-$}}
\put(50,90){\circle*{5}}
\put(46,21) {\ft{$\ $}}
\put(35,49) {$\mu$}
\put(50,21){\circle*{5}}
\put(10,20) {\fpe{$\mu^+$}}
\put(50,90) {\hpe{$\ $}{20}}
\put(52,21) {\bp{$\ $}{30}}
\put(85,93) {$A^0$}
\put(85,29) {$Z$}
\put(117,49){\makebox{\quad +\quad crossed diagram}}
\put(25,-14) {{\bf Figure.3}}
\end{picture}
\end{center}
\vspace{1.2cm}

The mechanism $\mu^+\mu^-\to A^0Z$ would provide an alternative way of
searching for $A^0$
whose detection is not guaranteed at the LHC or a $\sqrt s=500$
GeV $e^+e^-$ collider. At the latter this is because the conventional
production mechanism $e^+e^- \to
Z^* \to A^0H^0$ would be closed kinematically for
$M_A\approx M_H\ge 250$ GeV,
and $e^+e^- \to Z^* \to A^0h^0$ ($\sim \cos^2(\beta-\alpha)$)
is strongly suppressed for $M_A \ge 200$ GeV.
The proposed search at a $\mu^+\mu^-$ collider for
$M_A\ge \sqrt s/2$ is by doing a scan over $\sqrt s$ energies, in order
to find a resonance at $\sqrt s=M_A$, or by running the collider
at full $\sqrt s$ and looking for peaks in the $b\overline b$ mass
distribution (Bremsstrahlung tail method).
These methods are competitive and both may allow detection up to
$M_A\approx \sqrt s$ as long as $\tan\beta\ge 4-6$. However, both may
provide
quite a demanding challenge for detector resolution and machine design
(see Ref. \cite{Muon}),  and it is too early to say with certainty
if they would be feasible
methods in practice. With this is mind we consider the process
$\mu^+\mu^-\to A^0Z$.
With a sizeable rate for $\sigma(\mu^+\mu^- \to A Z)$,
$A^0$ could be discovered first in this channel,
and then the beams could be adjusted to $\sqrt s=M_A$
for precision studies. In addition, $\mu^+\mu^-\to A^0Z$
probes greater masses of $M_A$ than  $e^+e^-\to Z^* \to A^0H^0$,
and  becomes another option to first discover $A^0$ (if
discovery has been elusive at the LHC or a $\sqrt s=500$ GeV
$e^+e^-$ collider).
The matrix element squared may be written as:
\begin{eqnarray}
|{\cal M}|^2 &=& \frac{ sg^4m_{\mu}^2}{32M_W^4 } [
|a_V|^2 \lambda(s,M_A^2,M_Z^2)
-2a_{t1}g_A(M_A^2M_Z^2-sp_T^2-t^2)\Re(a_V) \nonumber \\ [0.15cm] & &
-2a_{t2}g_A(M_A^2M_Z^2-sp_T^2-u^2)\Re(a_V)\nonumber \\ [0.15cm] & &
+(g_A^2+g_V^2) \left\{
a_{t1}^2(2M_Z^2p_T^2+t^2)+a_{t2}^2(2M_Z^2p_T^2+u^2) \right\} \nonumber \\[0.15cm]
& &  -2(g_A^2-g_V^2)a_{t1}a_{t2}(2M_Z^2p_T^2+2M_A^2M_Z^2-tu) ]
\end{eqnarray}
with $a_V$ the same as in Section 2.2 and
\begin{eqnarray}
a_{t1} &=& {\lambda_{A\mu^+\mu^-}\over t-m_{\mu}^2} \qquad , \qquad
a_{t2} = {\lambda_{A\mu^+\mu^-}\over u-m_{\mu}^2}
\end{eqnarray}
The differential cross--section follows from eq(\ref{ref1}) with the changes
$M_{H^{\pm}}\to M_A$ and $M_W\to M_Z$.\\ 
The total cross--section is given by:
\begin{eqnarray}
\sigma(\mu^+\mu^- \to  A^0 Z)& = & \frac{G_F m_\mu^2}{32 \pi s^2 }
\{ [4 s (g_A^2-g_V^2) \tan^2\beta  +
 2 s |a_V|^2 \lambda (s,M_A^2,M_Z^2) ]
\lambda^{\frac{1}{2}}(s,M_A^2,M_Z^2)   \nonumber \\ [0.11cm] & + &
[ 8 s \Re(a_V) g_A (M_A^2 + M_Z^2 - s) \tan\beta  -
  8  (g_A^2 - g_V^2)   M_Z^2 \tan^2\beta  \nonumber \\ [0.11cm]& + &
  4  (g_A^2+ g_V^2)   (s-4 M_Z^2) \tan^2\beta  ] \lambda^{\frac{1}{2}}
(s,M_A^2,M_Z^2)  \\ & +  &
\frac{F(s,M_A^2,M_Z^2)}{(M_A^2 + M_Z^2 - s) }
[ -8 M_Z^2 \tan\beta ( -2  \Re(a_V)  g_A M_A^2 (M_A^2 + M_Z^2 - s) s
\nonumber \\ [0.11cm] & + &
    ( 2 (g_A^2-g_V^2) M_A^2 (M_Z^2 - s) +
  (g_A^2+g_V^2) (M_A^2 + M_Z^2 - s)^2 ) \tan\beta)     ] \} \nonumber
\end{eqnarray}

\section{Numerical results}
We now present our numerical analysis in the context of the MSSM.
We take $\sqrt s=500$ GeV and assume integrated luminosities of the order
50 fb$^{-1}$.

In Fig.~4
we plot $\sigma(\mu^+\mu^-\to H^{\pm}W^{\mp})$, defined as the sum of
$H^+W^-$ and $H^-W^+$ production, as a function of $M_{H^{\pm}}$,
varying $\tan\beta$ from 20 to 50. We also include the tree--level rate for
$\sigma(e^+e^-\to H^+H^-)$
in order to show the advantage of a $\mu^+\mu^-$ collider over an
$e^+e^-$ collider. One can see that the single production mode gains
in importance with increasing $\tan\beta$, and offers detection possibilities
for $M_{H^{\pm}}$ up to $\sqrt s - M_W$. This compares favourably with the
reach at an $e^+e^-$ collider.

The slight dip and rise of the
curves arises due to the $H^0$ and $A^0$ mediated s--channel contributions
increasing in magnitude with $M_{H^{\pm}}$, which compensates for the
phase space suppression until the kinematical limit is approached.
This can be seen from the fact that since $M_{H^{\pm}}\approx M_H\approx M_A$,
larger $M_{H^{\pm}}$ causes both $M_H$ and $M_A$ to be closer to
$\sqrt s$ (i.e. the resonance condition).

\setcounter{figure}{3}
\begin{figure}[htb]
\centerline{\protect\hbox{\psfig{file=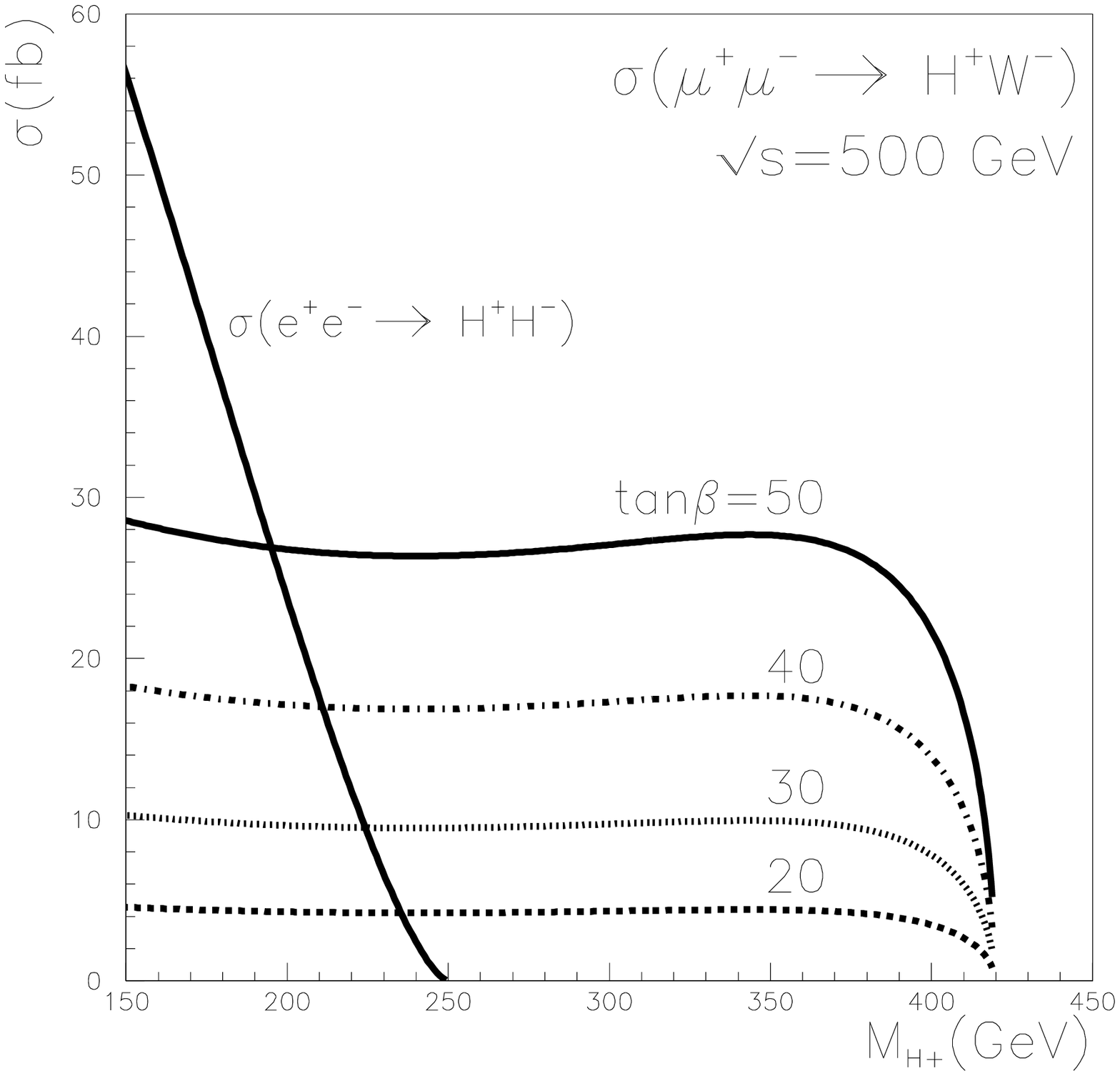,height=10cm,width=10cm}}}
\caption{$\sigma(\mu^+\mu^-\to H^{\pm}W^{\mp})$ as a function of
$M_{H^{\pm}}$ for various values of $\tan\beta$.
Also indicated is $\sigma(e^+e^-\to H^+H^-)$ for the same $\sqrt s$.}
\label{cha6}
\end{figure}
It is clear from the graphs that for $\tan\beta\ge 20$
one has $\sigma(\mu^+\mu^-\to H^{\pm}W^{\mp})\ge 5$ fb, which would
give a sizeable number of singly produced $H^{\pm}$ for
luminosities of 50 fb$^{-1}$.
One would expect $H^{\pm}\to tb$ decays for the mass region of interest
and so the main background would be from $t\overline t$ production.
Such a background \cite{MorOda} was shown to overwhelm the channel
$pp\to H^{\pm}W^{\mp}$ \cite{bbwh} at the LHC. However, at a $\sqrt s=500$ GeV
muon collider $\sigma(\mu^+\mu^-\to t\overline t)\sim 0.7$ pb in contrast to
$\sim 800$ pb at the LHC. Hence we would expect much better prospects for
detection at a muon collider although a full signal--background analysis
is beyond the scope of this paper. Previous studies of backgrounds to
$H^{\pm}W^{\mp}$ production at $e^+e^-$ colliders have been carried out
in the context of Higgs triplet models \cite{Phil}, assuming
$H^{\pm}\to W^{\pm}Z$ as
the main decay channel. Such studies cannot be applied to the MSSM
where $H^{\pm}\to tb$ decays would dominate.

We note that a 2HDM with the Model I type structure would not register
an observable signal in this channel. This is due to the rate being
proportional to $\cot^2\beta$, and so unacceptably small values of $\tan\beta$
would be required in order to allow observable cross--sections.

The process $\mu^+\mu^-\to A^0Z$ suffers from smaller
cross--sections and these are plotted as a function of $M_A$ in Fig.~5.
Given that $\mu^+\mu^-\to A^0H^0$ probes $M_A$ up to $\approx \sqrt s/2$
the region $M_A\ge 250$ GeV is of interest. We see that cross--sections
$\ge 1$ fb are only attainable in this region for $\tan\beta\ge 30$ and so
detection would be restricted to large values of $\tan\beta$. The smallness
of the cross--sections is caused by large destructive interference between
the $s$ and $t$ channels.

Finally, we consider $\mu^+\mu^-\to H^+H^-$. We find very small deviations
from the rate for $e^+e^-\to H^+H^-$, of the order a few percent for
large values of $\tan\beta$. This can be traced to the fact that the
s--channel Higgs exchange diagrams are far from resonance, and the t--channel
diagrams are doubly Yukawa suppressed. Since the  1--loop corrections
\cite{eeHH} may be much larger than these deviations we do not plot a graph.

\begin{figure}[hbt]
\centerline{\protect\hbox{\psfig{file=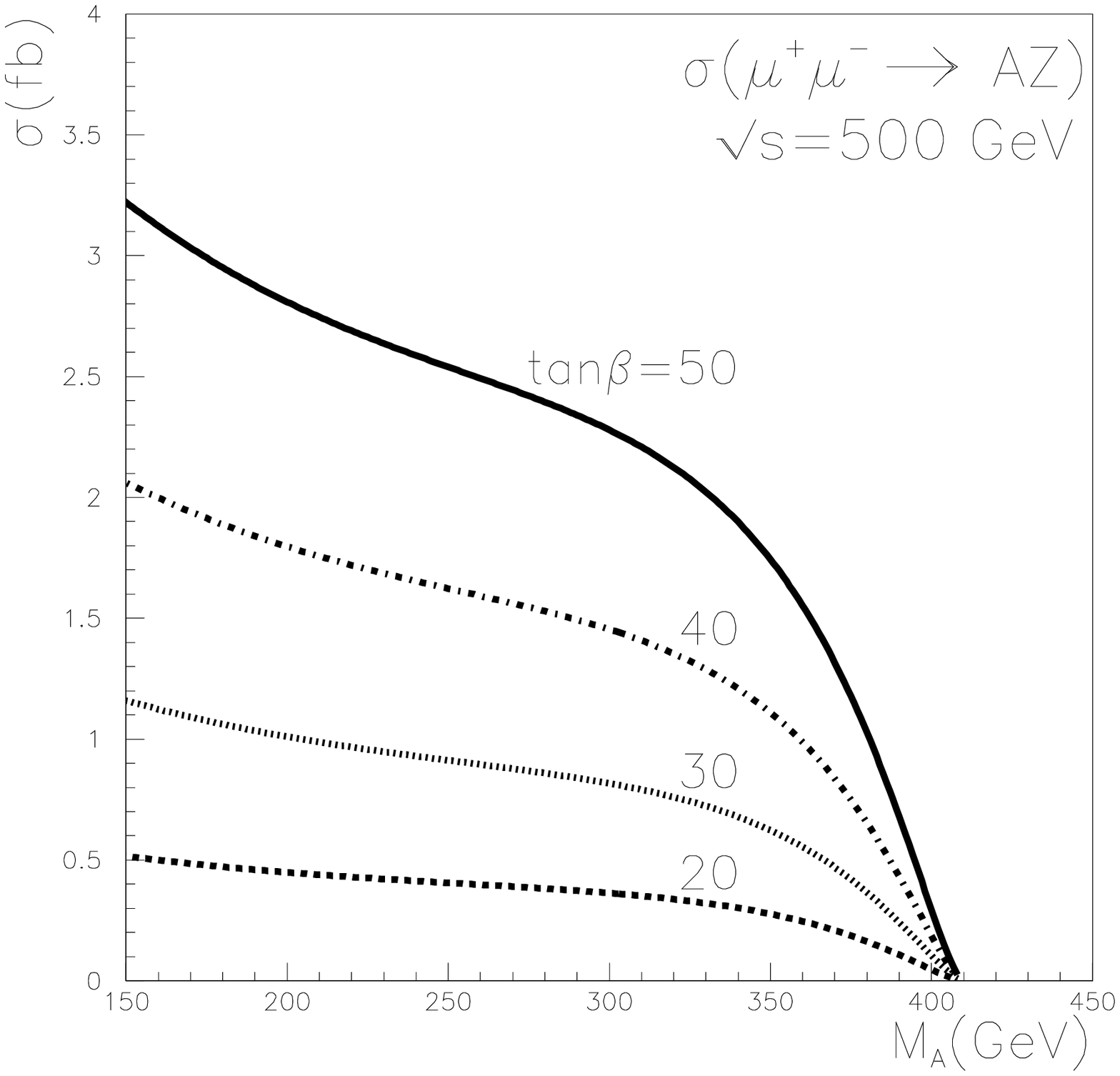,height=10cm,width=10cm}}}
\caption{$\sigma(\mu^+\mu^-\to A^0Z)$ as a function of $M_A$
for various values of $\tan\beta$.}
\label{cha6}
\end{figure}
\section{Conclusions}
We have considered the processes $\mu^+\mu^-\to H^{\pm}W^{\mp}$
and $\mu^+\mu^-\to A^0Z$ of the MSSM in the context of a
high--energy $\mu^+\mu^-$
collider ($\sqrt s=500$ GeV). We showed that $\mu^+\mu^-\to H^{\pm}W^{\mp}$
production offers an attractive new way of searching for $H^{\pm}$ at
such colliders. The cross--section grows with increasing $\tan\beta$
with values as large as 30 fb being attainable for $\tan\beta\ge 50$.
With an integrated luminosity of 50 fb$^{-1}$ a significant number
of $H^{\pm}$ could be produced singly up to $M_{H^{\pm}}\approx \sqrt s-M_W$.
This compares favourably with the reach at an $e^+e^-$ collider, which may only
probe up to $M_{H^{\pm}}\approx \sqrt s/2$.
The main background (assuming
$H^{\pm}\to tb$ decays) would be from $t\overline t$ production, which has
a cross--section of 700 fb, 3 orders of magnitude less than at the LHC.
We conclude that the mechanism  $\mu^+\mu^-\to H^{\pm}W^{\mp}$ represents
a novel and attractive way of producing $H^{\pm}$ at a $\mu^+\mu^-$ collider,
and in our opinion merits a detailed signal--background analysis.

Pseudoscalar Higgs production via $\mu^+\mu^-\to A^0Z$ offers smaller
cross--sections, with values of 2 fb or more only possible for large
($\ge 40$) $\tan\beta$. Charged Higgs pair production has essentially the
same rate as that at an $e^+e^-$ collider, with differences of the order
of a few percent for large values of $\tan\beta$.

\section*{Acknowledgements}
A.G. Akeroyd was supported by the Japan Society for Promotion of Science (JSPS).
We thank A. Turcot for useful comments.

\end{document}